\documentclass[aps,prl,twocolumn,superscriptaddress,showpacs,floatfix,nofootinbib]{revtex4}
\usepackage{amsmath,graphicx}

\newcommand{\mean}[1]{\left\langle #1 \right\rangle}
\newcommand{\smean}[1]{\langle #1 \rangle}
\newcommand{\N}{N_{\rm evts}}

\begin{document}

\preprint{Saclay-T03/097,TIFR/TH/03-...}

\title{Genuine collective flow from Lee-Yang zeroes}

\author{R. S. Bhalerao}
\affiliation{Department of Theoretical Physics, Tata Institute
of Fundamental Research, Homi Bhabha Road, Colaba, Mumbai 400 005,
India}

\author{N. Borghini}

\author{J.-Y. Ollitrault}
\affiliation{Service de Physique Th{\'e}orique, CEA-Saclay,
F-91191 Gif-sur-Yvette cedex, France}

\date{\today}

\begin{abstract}
We propose to use Lee-Yang theory of phase transitions as a practical tool 
to analyze experimentally anisotropic flow in nucleus-nucleus collisions. 
We argue that this method is more reliable than any other method, 
and that it is the natural way to analyze collective effects. 
\end{abstract}


\maketitle
Fifty years ago, Yang and Lee~\cite{Yang:be} showed that 
phase transitions can be characterized by the locations of the 
zeroes of the grand partition function in the complex plane. 
Since then, their theory has been extensively used, in particular, 
to study phase transitions in finite-size systems, via numerical 
simulations~\cite{Chen}: in lattice calculations, it has been applied 
to the electroweak~\cite{Csikor:1998eu} and QCD phase 
transitions~\cite{Fodor:2001au}.

\begin{center}
\begin{figure}[ht!]
\centerline{\includegraphics*[width=0.6\linewidth]{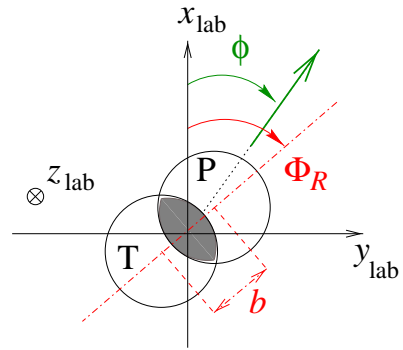}}
\caption{Schematic picture of a nucleus-nucleus 
collision viewed in the plane transverse to the collision axis $z$.
$b$ is the impact parameter, $\Phi_R$ its azimuthal angle. 
$\phi$ is the azimuthal angle of an outgoing particle.}
\label{fig:collision}
\end{figure}
\end{center}

In this Letter, we propose to apply Lee-Yang theory for the first
time to the analysis of experimental 
data.\footnote{Lee-Yang zeroes were already used in analyzing 
multiplicity distributions in high-energy collisions.
But it was shown that the locations of the zeroes merely reflect general, 
well-known features of these distributions~\cite{Brooks:1997kd}, 
and do not bring any new insight into the reaction dynamics.}
More specifically, we show that it is the most natural way to 
study anisotropic flow in nucleus-nucleus collisions. 
Anisotropic flow is defined as a correlation between the azimuthal 
angle $\phi$ of an outgoing particle and the azimuthal angle $\Phi_R$ 
of the impact parameter (see Fig.~\ref{fig:collision};
$\Phi_R$ is also called the orientation of the reaction plane), 
which is best characterized by the Fourier 
coefficients of the single-particle distribution~\cite{Voloshin:1996mz}:
\begin{equation}
\label{defivn}
v_n\equiv \mean{\cos n(\phi-\Phi_R)}.
\end{equation}
In this expression, $n$ is a positive integer and angular brackets 
denote an average over many particles belonging to some phase-space 
region, and over many collisions having approximately the same 
impact parameter. 
In particular, the so-called elliptic flow~\cite{Ollitrault:bk} 
$v_2$ is a sensitive probe of the 
dense matter produced in a nucleus-nucleus collision
at ultrarelativistic energies~\cite{Ackermann:2000tr}.

While $v_n$, defined by Eq.~(\ref{defivn}), is a trivial one-particle 
observable which can easily 
be computed in a model or an event generator, the experimental 
situation is quite different. Indeed, the reference direction 
$\Phi_R$ is unknown experimentally, and $v_n$ can only 
be measured indirectly, from the azimuthal correlations between 
the detected particles. Furthermore, $\Phi_R$ varies randomly from one 
event to the other, which has a remarkable consequence:
anisotropic flow appears as a truly collective motion, in the sense
that all outgoing particles in a given event seem to be attracted 
towards some arbitrary direction. 

The standard method for analyzing anisotropic flow 
is to correlate particles with an estimate of 
$\Phi_R$~\cite{Danielewicz:hn}. However, this 
estimate is itself obtained from the outgoing particles, 
and one essentially measures a two-particle correlation~\cite{Wang:1991qh}. 
Intuitively, two-body correlations are not the appropriate tool 
to probe collective behaviour. 
Indeed, these two-particle methods were shown to be inadequate 
due to various ``nonflow'' correlations from 
quantum statistics~\cite{Dinh:1999mn}, resonance decays, 
minijet production~\cite{Kovchegov:2002nf}, etc., 
which are neglected and bias the analysis. 
Recently, new methods were developed, based on higher-order 
(typically, four-particle) correlations, together 
with a cumulant expansion which eliminates low-order nonflow
correlations~\cite{Borghini:2000sa}. 
However, it was argued that experimental results~\cite{Adler:2002pu}
could still be biased by nonflow effects~\cite{Kovchegov:2002cd} 
at this order.
In this paper, $v_n$ will be analyzed directly from 
the correlation between a large number of particles. 
It will be shown that the results are perfectly stable
with respect to nonflow correlations, which involve a 
smaller number of particles. 

Our new method is based on the following global observable, which 
is defined for each event:
\begin{equation}
\label{defqtheta}
Q^\theta= \sum_{j=1}^M \cos n(\phi_j-\theta),
\end{equation}
where $n$ is the Fourier harmonic under study ($n=1$ for directed
flow $v_1$, $n=2$ for elliptic flow), the sum runs over all 
$M$ detected particles, $\phi_j$ are their azimuthal angles, 
and $\theta$ is an arbitrary reference direction. 
This quantity is nothing but a projection of the ``event flow-vector,'' 
used in other methods to estimate the orientation of the reaction 
plane~\cite{Danielewicz:hn}, on the transverse direction making an 
angle $n \theta$ with respect to the $x$-axis. 
In practice, the sum in Eq.~(\ref{defqtheta}) is often weighted:
weights depending on the particle mass, transverse momentum and rapidity 
are used in order to reduce statistical errors and increase the flow signal. 
They are omitted here for the sake of simplicity, but should be included 
in the actual analysis. 

The central object in the method is 
the moment generating function~\cite{vanKampen}
\begin{equation}
\label{defGint}
G(z)\equiv \mean{e^{z Q^\theta}},
\end{equation}
where $z$ is a complex variable, and 
angular brackets now denote an average over a large number of 
events with the same impact parameter. 
The procedure to obtain $v_n$ (as will be shown below) is the following:
choose a value of $\theta$; construct $Q^\theta$ for each 
event, evaluate $G({\rm i}r)$ for real, positive $r$; plot 
$|G({\rm i}r)|$ as a function of $r$; determine the first minimum 
$r_0$. The flow is given by $v_n\simeq 2.405/Mr_0$.

Let us now justify the procedure.
We first introduce the cumulants $c_k$, which are defined 
as~\cite{vanKampen} 
\begin{equation}
\label{defcumulint}
\ln G(z)\equiv\sum_{k=1}^{+\infty} \frac{c_k}{k!} z^k.
\end{equation}
The first two terms in this power-series expansion correspond to 
the average value of $Q^\theta$, and the square of its standard
deviation, respectively:
\begin{equation}
\label{c12}
c_1 = \mean{Q^\theta},\ \ \ c_2 = \mean{(Q^\theta)^2}-\mean{Q^\theta}^2.
\end{equation}
Note that $c_1$ vanishes by symmetry if the detector
has uniform azimuthal coverage. 

The order of magnitude of the cumulants differs depending
on whether or not there are collective effects in the system.
Since $Q^\theta$ is the sum of $M$ terms of order unity, 
and $c_k$ involves $(Q^\theta)^k$, the naive expectation is 
that $c_k$ should be of order  $M^k$ or, more generally, scale 
with $M$ like $M^k$. 
As we shall see later, this is precisely the case when 
anisotropic flow is present. 
When no collective effect is present, however, cumulants 
are much smaller: one can view the system as made of 
independent clusters of particles.
$G(z)$ then factorizes into the product of the contributions 
of each cluster, which is converted into a sum by the logarithm. 
Hence $c_k$ scales only {\it linearly\/} with $M$.
For particles emitted with uncorrelated, randomly distributed  
azimuthal angles, for instance, Eq.~(\ref{c12}) gives $c_2=M/2$. 

This shows that the value of $c_k$ for large $k$ is the natural 
observable to characterize collective effects:
the larger $k$, the larger the contribution of collective effects, 
relative to other contributions, which scale like $M$.
The asymptotic behaviour of $c_k$ for large $k$ therefore provides 
a cleaner separation between collective effects and few-body 
correlations than finite-order cumulants~\cite{Borghini:2000sa}. 
This can be easily understood physically: 
the cumulant  $c_k$ essentially isolates the contribution of 
genuine $k$-particle correlations, by subtracting out most 
contributions from lower-order correlations.  
In order to study collective effects, which by 
definition involve a large number of particles, $k$ should be 
as large as possible. 

The asymptotic behaviour of $c_k$ for large $k$ is determined by the radius 
of convergence of the power-series expansion, Eq.~(\ref{defcumulint}), 
i.e., by the singularity of $\ln G(z)$ which lies closest to the 
origin in the complex plane. 
Since $G(z)$ has no singularity, the only possible singularities
of $\ln G(z)$ are the zeroes of $G(z)$. 
If $z_0$ denotes the zero closest to the origin, $c_k$ 
scales typically like $z_0^{-k}$ for large $k$. 
Therefore, if $c_k$ scales like $M^k$ (collective effects), 
$z_0$ scales like $1/M$. If there is no collective effect, 
$G(z)$ is the product of contributions of small clusters, 
and the zeroes of $G(z)$ are the zeroes of the individual
contributions:
$z_0$ does not depend on $M$. 

We are now in a position to explain how our approach relates 
to the theory of phase transitions of Yang and Lee~\cite{Yang:be}. 
The starting point is the grand partition function:
\begin{equation}
\label{zmu}
{\cal G}(\mu)=\sum_{N=0}^{+\infty} Z_N e^{\mu N/kT},
\end{equation}
where $Z_N$ is the canonical partition function for $N$ particles
at temperature $T$ in a volume $V$ (both $T$ and $V$ are fixed). 
Let $\mu_c$ denote a reference value of the chemical potential $\mu$.
The probability $P_N$ to have $N$ particles in the system at
$\mu=\mu_c$ is
\begin{equation}
\label{PN}
P_N\equiv\frac{Z_N e^{\mu_cN/kT}}{{\cal G}(\mu_c)}.
\end{equation}
The moment generating function of this probability 
distribution can be simply expressed in terms 
of the grand partition function, Eq.~(\ref{zmu})
\begin{equation}
\label{Gleeyang}
G(z)\equiv \sum_{N=0}^{+\infty} P_N e^{zN}
=\frac{{\cal G}(\mu_c+kTz)}{{\cal G}(\mu_c)}.
\end{equation}
This function is  analogous to our generating function, Eq.~(\ref{defGint}), 
with the number of particles $N$ instead of $Q^\theta$, 
and the volume $V$ instead of the multiplicity $M$. 
We can repeat the previous discussions:
if particles are correlated only within small clusters, 
$z_0$ (the zero of $G(z)$ closest to the origin) is 
independent of $V$. 
This is the case when no phase transition occurs at $\mu=\mu_c$. 
Now assume that a first-order transition, say, a liquid-gas
transition,  occurs at $\mu=\mu_c$. 
Then, the system can be any mixture of the low-density gas phase and
the high-density liquid phase. 
The probability distribution $P_N$ in Eq.~(\ref{Gleeyang}) 
is widely spread between two values $N_{\rm min}$ (gas) and $N_{\rm max}$
(liquid) which both scale like the volume $V$.
Then, the partition function $G(z)$ depends on the volume
$V$ essentially through the combination $zV$, and consequently its
zeroes scale with the volume like $1/V$.
The general result of Lee and Yang is precisely that a phase
transition occurs at $\mu=\mu_c$ if the zeroes of $G(z)$ come closer 
and closer to the origin $z=0$ as the volume of the system, $V$, 
increases 
(note, however, that Ref.~\cite{Yang:be} is written in terms of 
the variable $y=e^z$ instead of $z$).

Let us come back to heavy ion collisions. So far, our analysis 
has been general, and  $Q^\theta$ could be replaced by 
any extensive variable in Eq.~(\ref{defGint}). 
We are now going to specify what happens when there is 
anisotropic flow in the system. 
We can repeat the discussion of Eqs.~(\ref{defGint}-\ref{c12}), 
but with all average values taken for a fixed orientation 
of the reaction plane $\Phi_R$. Such averages will be denoted by 
$\mean{\cdots|\Phi_R}$. 
Using the definition of $v_n$, Eq.~(\ref{defivn}),
and symmetry with respect to the reaction plane
(which implies $\smean{\sin n(\phi-\Phi_R)}=0$), 
and assuming for simplicity that the multiplicity $M$ 
is the same for all events, one obtains from Eq.~(\ref{defqtheta}):
\begin{equation}
\label{averageQ}
c_1=\mean{Q^\theta|\Phi_R}=M v_n\,\cos(n(\Phi_R-\theta)).
\end{equation}
We neglect terms $c_3$ and higher in Eq.~(\ref{defcumulint}). 
This amounts to assuming that the probability distribution of 
$Q^\theta$ is gaussian for a fixed $\Phi_R$: this is the central
limit theorem, which holds if $M$ is large enough, and if 
there is no other collective effect in the system. 
We further neglect the $\Phi_R$-dependence of $c_2$. 
Finally, averaging over $\Phi_R$, one obtains the following 
theoretical expression of $G(z)$, 
which we denote by $G_{\rm c. l.}(z)$ since it corresponds 
to the central limit approximation:
\begin{equation}
\label{meanGflowbis}
G_{\rm c.l.}(z)=e^{c_2 z^2/2}\, I_0(Mv_n z),
\end{equation}
where $I_0$ is a modified Bessel function.
Taking the logarithm and expanding in powers of $z$, one checks
that the cumulant $c_k$ in Eq.(\ref{defcumulint}) scales with 
$M$ like $M^k$, as anticipated. 

The first zeroes of $G_{\rm c.l.}(z)$ lie on the imaginary axis at 
\begin{equation}
\label{defz0}
z_0={\rm i}r_0=\frac{{\rm i}j_{01}}{Mv_n},
\end{equation}
and at $-z_0$, 
where $j_{01}\simeq 2.405$ is the first positive root of the Bessel function 
$J_0(x)$. As expected from the general discussion above, anisotropic 
flow $v_n$, being a collective effect, is completely determined by
$z_0$. 
The situation is analogous to a first-order phase transition, in the sense 
that the position of the zero scales like $1/M$, and the multiplicity 
$M$ is the analogue of the volume $V$ in Lee-Yang's theory. 
The important difference with statistical physics 
is that the system size is much smaller. 
As a consequence, zeroes never come very close to the
origin, but the physics involved is essentially
the same.

In a second paper~\cite{Lee:1952ig}, Lee and Yang further showed 
that all zeroes lie on the imaginary axis of the variable $z$
(or, equivalently, on the unit circle for $y=e^z$) for a 
general class of models. It is interesting to note that 
our theoretical estimate, Eq.~(\ref{meanGflowbis}), has the 
same property.

\begin{center}
\begin{figure}[ht!]
\centerline{\includegraphics*[width=\linewidth]{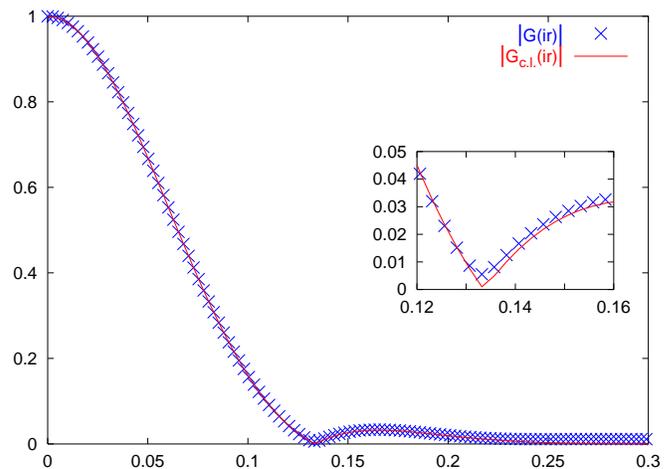}}
\caption{Variation of $\left|G(ir)\right|$ with $r$.
The crosses are the values of $|G({\rm i}r)|$. 
The solid line displays the expected value $|G_{\rm c.l.}({\rm i}r)|$
defined by Eq.~(\ref{meanGflowbis}).}
\label{fig:Gvsr}
\end{figure}
\end{center}

Figure~\ref{fig:Gvsr} displays the variation of 
$\left|G({\rm i}r)\right|$ as a function of $r$ for simulated data. 
The data set contained $\N=20000$ events. In each 
event, $M=300$ particles are emitted independently with an azimuthal 
distribution $dN/d\phi\propto 1+2 v_2\cos(2(\phi-\Phi_R))$, 
where $v_2=6\%$, and the azimuth of the reaction plane, $\Phi_R$, 
is randomly chosen. 
These numbers are typical values for a mid-central Au+Au collision at 
$\sqrt{s_{_{\rm NN}}} = 130$~GeV, as analyzed by the STAR 
Collaboration~\cite{Ackermann:2000tr}.
The global observable $Q^\theta$ in  Eq.~(\ref{defqtheta}) was 
constructed for each event with $n=2$ and various values of $\theta$. 
Figure~\ref{fig:Gvsr} corresponds to $\theta=0$. 
The numerical results are compared with the theoretical estimate, 
Eq.~(\ref{meanGflowbis}), where we have taken $c_2=M/2$ 
(the expected value for independent particles). The excellent agreement 
justifies the approximations made in deriving Eq.~(\ref{meanGflowbis}).
However, a closer look at the numerical results 
(inlay in Fig.~\ref{fig:Gvsr}) shows that unlike 
the theoretical estimate, $|G({\rm i}r)|$ does not strictly vanish:
due to statistical fluctuations, the zeroes of $G(z)$ are 
slightly off the imaginary axis. 
This small deviation is physically irrelevant, and we 
choose to investigate the minima of $|G(z)|$, 
rather than the zeroes of $G(z)$. 
We denote by $r_0^\theta$ the first minimum of 
$|G({\rm i}r)|$, where the superscript $\theta$ recalls that it
may depend on the reference angle $\theta$ in Eq.~(\ref{defqtheta}). 
Identifying $G(z)$ with the theoretical estimate $G_{\rm c.l.}(z)$, 
and using Eq.~(\ref{defz0}), we obtain the following estimate 
of $v_n$, which may also depend on $\theta$:
\begin{equation}
\label{flowestimate0}
v_n^\theta\equiv\frac{j_{01}}{Mr_0^\theta}.
\end{equation}

\begin{center}
\begin{figure}[ht!]
\centerline{\includegraphics*[width=\linewidth]{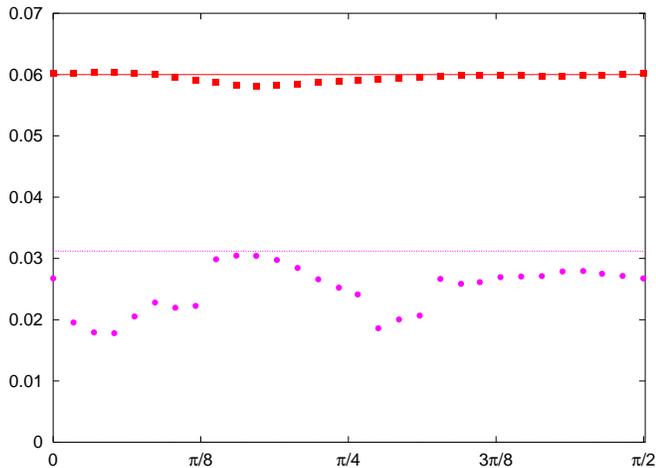}}
\caption{The reconstructed value $v_2^\theta$ 
as a function of $\theta$. 
Squares: simulated data with input $v_2=6\%$
(same data as in Fig.~\ref{fig:Gvsr}). 
Circles: simulated data with input $v_2=0$. 
}
\label{fig:plotv}
\end{figure}
\end{center}
This procedure was applied to the simulated data. 
The result is shown in Fig.~\ref{fig:plotv}. 
$v_2^\theta$ coincides with the input value $v_2=6\%$, up to 
statistical fluctuations. Performing the analysis for several values
of $\theta$, and averaging $v_2^\theta$ over $\theta$, 
reduces this statistical error. We finally obtain  $5.95\%$:
$v_2$ is reconstructed with great accuracy.

For the sake of illustration, we also applied the same procedure
to simulated data with no flow. The procedure yields a spurious flow, 
due to statistical fluctuations, which is also shown 
in Fig.~\ref{fig:plotv}. 
The magnitude of this spurious flow can easily be understood.
The average in Eq.~(\ref{defGint}) is evaluated over a finite 
number of events, $\N$. 
As a consequence, $G({\rm i}r)$ has statistical fluctuations, whose 
typical magnitude is $1/\sqrt{\N}$. 
For large enough $r$, they become as large as the expectation 
value given by Eq.~(\ref{meanGflowbis}), in which we set $v_n=0$
and $c_2=M/2$. This occurs when 
\begin{equation}
e^{-Mr^2/4}\sim \frac{1}{\sqrt{\N}}.
\end{equation}
As soon as $r$ is larger than this value, fluctuations 
can produce a minimum of $|G({\rm i}r)|$. 
The corresponding ``spurious flow'' given by the analysis, 
Eq.~(\ref{flowestimate0}), satisfies 
\begin{equation}
\label{spurious}
v_n^\theta\lesssim\frac{j_{01}}{\sqrt{2 M\ln\N}}.
\end{equation}
For our simulated data, the right-hand side (rhs) is about $3.1\%$, which 
is depicted as the dashed line in Fig.~\ref{fig:plotv}. 
As expected, the values of $v_2^\theta$ lie below this value, 
but only slightly. 
This is the main limitation of our method: $v_n$ can be 
safely reconstructed only if it is larger than the rhs of 
Eq.~(\ref{spurious}).

Since no ``nonflow'' correlation between the particles
was simulated, standard methods of flow analysis would have 
worked well too. However, the unique feature of the present 
method is its absolute stability with respect 
to such correlations. As an illustration,
assume that instead of emitting $M$ particles in each event, 
we emit $M$ clusters, each cluster containing $q$ collinear particles. 
Then, $Q^\theta$ is increased by a factor of $q$. 
As a consequence, the position of the first minimum, $r_0^\theta$, 
is smaller by a factor of $q$. Since the event multiplicity
is now $qM$, one must replace $M$ with $qM$ in the denominator
of Eq.~(\ref{flowestimate0}), so that the flow estimate 
$v_n^\theta$ is strictly the same, as it should. 
On the other hand, estimates of $v_2$ from 2-particle or 4-particle 
methods~\cite{Borghini:2000sa} are generally increased by such 
correlations.
With the numerical values above, the increase would be 
significant for 2-particle methods (one obtains $v_2=7.3\%$ 
instead of $6\%$), but very small for 4-particle cumulants: 
in most cases of interest, these cumulants will give results very 
similar to those obtained with the present method, 
but the latter is the most systematic one to disentangle 
collective motion from other effects.

The method can be extended to the analysis of differential flow, 
i.e., the analysis of $v_n$ as a function of transverse momentum 
and rapidity. This is explained in detail in 
Ref.~\cite{bbo}, where we also discuss in detail errors due to nonflow 
correlations, statistical fluctuations, and show that 
the method is remarkably insensitive to azimuthal asymmetries 
in the detector acceptance.

We have shown that Lee-Yang theory of phase transitions can be 
used as a practical means of analyzing anisotropic flow 
experimentally. The method 
is expected to give results similar to cumulant methods, but 
is significantly simpler to implement, and formally elegant. 
It does not require the knowledge of the reaction plane 
and there is no need to construct correlation functions
and cumulants. 
More generally, Lee-Yang zeroes provide a natural probe of 
collective behaviour. It would be interesting to extend the 
present approach to other observables, 
in order to look for critical fluctuations which may occur 
in the vicinity of a phase transition.~\cite{Jeon:2003gk}

\section*{Acknowledgments}

R.~S.~B. acknowledges the hospitality of the SPhT, CEA, Saclay;
J.-Y.~O. acknowledges the hospitality of the Department of Theoretical
Physics, TIFR, Mumbai.
Both acknowledge the financial support from CEFIPRA, New Delhi, 
under its project no. 2104-02.

\end{document}